\documentclass[aps,12pt,superscriptaddress,amsfonts,amssymb,amsmath]{revtex4}
\pdfoutput=1

\usepackage{graphicx}
\usepackage{epsfig}
\usepackage{makeidx}
\usepackage{epstopdf}
\usepackage{natbib}

\begin{document}

\title{Metrics with zero and almost-zero Einstein action \\ in quantum gravity
}

\author{G.\ Modanese \footnote{Email address: giovanni.modanese@unibz.it}}
\affiliation{Free University of Bolzano-Bozen \\ Faculty of Science and Technology \\ I-39100 Bolzano, Italy}

\linespread{0.9}

\begin{abstract}

We generate numerically on a lattice an ensemble of stationary metrics, with spherical symmetry, which have Einstein action $S_E \ll \hbar$. This is obtained through a Metropolis algorithm with weight $\exp(-\beta^2 S^2_E)$ and $\beta \gg \hbar^{-1}$. The squared action in the exponential allows to circumvent the problem of the non-positivity of $S_E$. The discretized metrics obtained exhibit a spontaneous polarization in regions of positive and negative scalar curvature. We compare this ensemble with a class of continuous metrics previously found, which satisfy the condition $S_E=0$ exactly, or in certain cases even the stronger condition $R({\bf x})=0$ for any ${\bf x}$. All these gravitational field configurations are of considerable interest in quantum gravity, because they represent possible vacuum fluctuations and are markedly different from Wheeler's ``spacetime foam''.

\end{abstract}

\maketitle

\section{Introduction}

Among the obstacles encountered in the formulation of quantum gravity theories one can certainly mention the difficulties in the Euclidean formulation (analytical continuation to imaginary time), which is usually employed in discretized lattice versions of quantum field theory \cite{zinn1996quantum,hamber2008quantum}. This problem is in turn related to the non-positivity of the Einstein action \cite{wetterich1998effective,modanese1998stability}.

The two main current approaches to lattice quantum gravity, namely those of Hamber and collaborators \cite{hamber2008quantum,hamber2009quantum,hamber2019vacuum} and Ambj\o rn and collaborators \cite{ambjorn2012nonperturbative} take a quite radical stance in the general re-definition of 4D spacetime as a discrete quantum dynamical object. Therefore the answers given in these approaches to the issues of stability and analyical continuation are not simple to translate into familiar quantum field theory language and can be possibly understood in ``entropic'' terms (see for instance the discussion of the conformal instability in \cite{hamber2008quantum}). 

The stability issue is also related, in our opinion, to one of the crucial questions in quantum gravity, particularly from a path integral point of view: among the infinite possible configurations of a void spacetime (many of which singular), why do we see on average a flat metric, and what are the most important fluctuations? Our aim in this work is not to give a general answer to this question, but only to obtain some insights in a simplified case, namely a path integral of the form 
\begin{align}
\int d[g_{\mu \nu}] \exp \left( \frac{i}{\hbar} S_E \right) =
	\int d[g_{\mu \nu}] \exp \left[ \frac{i}{\hbar} \left( -\frac{1}{16\pi G} \right) \int d^4x \sqrt{g(x)}R(x) \right]
	\label{eq1}
\end{align}
defined in the usual metric formalism and restricted to field configurations which are time-independent and spherically symmetric. Metrics for which $S_E=0$ or $S_E \ll \hbar$ clearly play an important role in this path integral.

In previous work we have been looking for stationary non-flat metrics, which we call ``zero modes of the Einstein action'', such that $\int d^4x\sqrt{g}R=0$. We found in \cite{modanese1999virtual,modanese2000large,modanese2000paradox} perturbative solutions (weak-field zero modes) with regions of opposite curvature, either in the form of dipoles or concentric shells. In \cite{modanese2007vacuum} we found exact non-perturbative solutions with spherical symmetry of the {\em local} condition $R=0$ and deformations of these solutions which still satisfy $\int d^4x\sqrt{g}R=0$, forming an infinite-dimensional functional space. 

The sets of these classical metrics can be regarded as extensions of the so-called Einstein spaces in the Petrov classification \cite{petrov1969einstein}, defined as the solutions of $R_{\mu \nu}=k g_{\mu \nu}$; by weakening this condition we can consider the condition $R_{\mu \nu}=0$, $R=0$ and finally $\int \sqrt{g}R=0$. While in General Relativity it is natural to devote special attention to the spaces with $R_{\mu \nu}=0$ (vacuum solutions of the Einstein equations), in a path integral for quantum gravity it is natural to focus on the condition $S_E=0$. Note that in usual field theories with a positive-definite action one never encounters such a distinction between the minima of the action and its zero-modes.

After finding these zero modes, however, we still do not know how much do they contribute to the path integral, unless we can evaluate it at least approximately. Working with the imaginary exponential $\exp(iS_E/\hbar)$ one can obtain some formal results \cite{modanese2012anomalous}, but a numerical lattice approach is not viable. If one tries to discretize the integral and perform it numerically on Wiener paths, one only obtains wild oscillations and a confirmation of the undefined sign of the action (Sect.\ \ref{sub}). 

So a new idea developed in this work is to explore the zero-modes on a lattice using a Metropolis algorithm of the same kind successfully employed in statistical mechanics based on the energy, with weight $\exp[-E/(k_BT)]$, or in Euclidean field theories with positive-defined actions. The method consists in starting from a metric which has zero action and is also a minimum of the action (flat space), then generate local deformations of this metric and accept them only if their action differs from zero by a quantity $\ll \hbar$, like in a semiclassical approximation of the Lorentzian path integral, but without any restriction to weak fields. For this purpose we use in the Metropolis algorithm the weight $\exp(-\beta^2 S^2_E)$, with $\beta \gg 1/\hbar$ (a limit which formally corresponds to a very low temperature). The choice of the squared action is the simplest one, although obviously not the only possible. Another simple choice is to use $|S_E|$ (Sect.\ \ref{alt}); the results are very similar, thus showing the robustness of the approach.

We start in Sect.\ \ref{sim} by laying out immediately our method for the calculation of the discretized action. Then in Sect.\ \ref{cla} we give a short review of previous work on classical, exact zero modes of the Einstein action. This serves as background and motivation for the present work. Sect.\ \ref{res} contains our results. In Sect.\ \ref{sim} we offer a comparison with other approaches to quantum gravity. Finally, Sect.\ \ref{con} contains our conclusions and an outlook on future work.

\section{Simplified recipe for the discretized path integral}
\label{sim}

We shall consider a specific subset of metric configurations, such that the lattice action takes a simple form. Even this subset contains, however, interesting candidates as unexpected vacuum fluctuations. 

\subsection{Spherically symmetric spaces with constant $g_{tt}$}
\label{sph}

Let us consider the set of all time-independent spherically symmetric metrics, and write the invariant interval as
\begin{align}
d\tau^2=-g_{\mu \nu}dx^\mu dx^\nu=B(r)dt^2-A(r)dr^2-r^2(d\theta^2+\sin^2\theta d\phi^2)
\end{align}
where $A(r)=g_{rr}(r)$ and $B(r)=g_{tt}(r)$ are two arbitrary smooth functions. The scalar curvature can then be expressed as \cite{weinberg1973gravitation}
\begin{align}
R=-\frac{R_{tt}}{B}+\frac{R_{rr}}{A}+2\frac{R_{\theta \theta}}{r^2}
\end{align}
where 
\begin{align}
R_{tt}=-\frac{B''}{2A}+\frac{B'}{4A}(a+b)-\frac{B'}{rA}
\end{align}
\begin{align}
	R_{rr}=\frac{B''}{2B}-\frac{1}{4}b(a+b)-\frac{a}{r}
\end{align}
\begin{align}
	R_{\theta \theta}=-1+\frac{r}{2A}(-a+b)+\frac{1}{A}
\end{align}
\begin{align}
a=\frac{A'}{A}; \ \ b=\frac{b'}{B}
\end{align}

If we limit ourselves to the simpler case $B=const.=1$, the entire expression for $R$ reduces to
\begin{align}
R=-\frac{2}{r}\left( \frac{A'}{A^2}+\frac{1}{r}-\frac{1}{Ar} \right)
\end{align}
When we multiply this by $\sqrt{g}$ and integrate in $d^3x$ and over a finite time interval $(-\tau,\tau)$, we obtain
\begin{align}
S_E=-\frac{1}{16\pi G}\cdot 2\tau \int_0^\infty dr \, 4\pi r^2 \cdot \frac{-2}{r}\sqrt{|A|}\left( \frac{A'}{A^2}+\frac{1}{r}-\frac{1}{Ar} \right)
\end{align}
\begin{align}
= \frac{\tau}{G} \int_0^\infty dr \, \sqrt{|A|} \left( \frac{rA'}{A^2}+1-\frac{1}{A} \right)
\label{eq8}
\end{align}

The path integral, reduced to this space, that we compute in the following, has the form
\begin{align}
\langle A(r_1) \rangle =\frac{\int d[A(r)]A(r_1)e^{iS_E/\hbar}}{\int d[A(r)]e^{iS/\hbar}},
\label{PI-A}
\end{align}
where $r_1 \in [0,+\infty)$ and we set the boundary condition $A(+\infty)=1$. For the moment we just trust that this makes sense and converges in spite of the oscillating factor $\exp(iS_E/\hbar)$, according to the original Feynman formulation (or better thanks to it). But we shall see that this is not numerically viable and it is better to turn to a properly modified Euclidean version.	

\subsection{Discretized action}

Let us choose units in which $G=c=\hbar=1$, and an integration time $\tau=1$.
We reduce the domain of $A(r)$ to a finite interval $(0,L)$ and divide it into $N$ parts, with $\delta=L/N$. Discretizing the action we obtain
\begin{align}
	S_E^{discr}=\delta \sum_{h=0}^N \sqrt{|\hat{A}_h|} \left( \frac{2h+1}{2\hat{A}_h^2} (A_{h+1}-A_h)+1-\frac{1}{\hat{A}_h} \right)
\end{align}
where $\hat{A}=(A_{h+1}+A_h)/2$. The initial conditions are $A_0,...,A_N=1$. The (fixed) boundary condition is $A_{N+1}=1$.

It is straightforward to see that if we change by $\pm \varepsilon$ one of the $A_h$ there are only two terms in the sum which are changed. The corresponding variation in $S_E^{discr}$ can be written explicitly, but it is safer to compute it numerically by difference at every Metropolis step.

\subsection{Comparison to the scalar field case and Lorentzian path integral}
\label{sub}

It is useful to make a comparison with a field $\phi(r)$ with classical vacuum value $\phi(r)=\phi_0$ and with the usual quadratic action
\begin{equation}
S_\phi=\int_0^\infty dr \left[ \left( \phi'(r) \right)^2+m^2\left( \phi(r)-\phi_0 \right)^2 \right]
\end{equation}
where we take in the following $m=1$, $\phi_0=1$, so that it has the same vacuum value as $A$.

Let us first try to compute numerically the Lorentzian path integral, even though we know from the start that there are little chances to obtain a convergent result.  We compute the quantity
\begin{equation}
\langle \phi_k \rangle = \frac{\int d[\phi] \phi_k e^{iS/\hbar}}{\int d[\phi] e^{iS/\hbar}}
\label{phik}
\end{equation}
This is not a real-valued average, but a complex ``transition element'', according to the Feynman denomination. As long as $S/\hbar \ll 1$ and $e^{iS/\hbar}\simeq 1$, we may expect that $\langle \phi_k \rangle$ is approximately real and close to 1. 

We write the discretized action as
\begin{equation}
S^{disc}_\phi = \delta \sum_{h=0}^N \left[ \left( \frac{\phi_{h+1}-\phi_h}{\delta} \right)^2+\left( \frac{\phi_{h+1}+\phi_h}{2}-1 \right)^2 \right]
\end{equation}
with the boundary condition $\phi_{N+1}=1$. 

Next we generate $N$ variables of the form  $\phi_h=1+\xi$, $-a<\xi<a$, where $\xi$ is a random number and the amplitude $a$ is initially very small (small deformations of the classical solution). The integrands in (\ref{phik}) are evaluated in correspondence of these values of $\phi_h$ and the procedure is repeated for a large number of times, averaging the results. In the limit of infinite $a$ this gives the so-called ``Wiener path integral'', actually an ordinary $N$-dimensional integral corresponding to a path integral along non-differentiable zigzag paths.

For small values of $a$ the fluctuations are small and we obtain for instance in the case $N=50$, $\hbar=1$ the reasonable numerical results of Tab.\ \ref{tab1}.

\begin{table}[h]
\begin{center}
\begin{tabular}{|c|c|c|c|}
\hline
$a$ & \ Re$\langle (\phi_{25}-1)^2 \rangle$ \ & \ Im$\langle (\phi_{25}-1)^2 \rangle$ \ & \ $S_{max}$ \ \\
\hline
$10^{-4}$ & $3.33 \cdot 10^{-9}$ & $2.78 \cdot 10^{-17}$ & $5\cdot 10^{-6}$  \\
$10^{-2}$ & $3.33 \cdot 10^{-5}$ & $2.78 \cdot 10^{-9}$ & $5\cdot 10^{-2}$  \\
1 & $(4\pm 2) \cdot 10^{-1}$ & $(1\pm 2) \cdot 10^{-1}$ & $5\cdot 10^2$  \\
\hline
\end{tabular}
\end{center}
\caption
{Feynman-Hibbs transition elements computed numerically in the Lorentzian path integral (\ref{phik}) in dependence on the amplitude $a$ of the integration interval. The last column gives the maximum value of the action recorded during the Monte Carlo integration, which runs typically over 10 to 100 million random values of the set $\phi_1,...,\phi_{50}$.}
\label{tab1}
\end{table}

The precision quickly decreases, however, when the amplitude $a$ increases, and for $a>1$ the real and imaginary parts of $\langle \phi_k \rangle$ undergo wild fluctuations even over very long runs. It appears that the ``destructive interference'' effect due to the oscillating factor $e^{iS/\hbar}$, which in principle should lead to the cancellation of the contributions from paths very far from the classical one, cannot in practice be reproduced numerically.

Nevertheless, since the action of the $\phi$-field is positive-defined, we can simulate a Lorentzian path integral with the limitation $S_\phi \ll \hbar$ as a thermal system at equilibrium at a suitable temperature $T$ using the Metropolis algorithm, i.e., starting from certain initial conditions (for instance the classical vacuum value $\phi_h=1$ for any $h$), making random changes of the $\phi_h$ and accepting them unconditionally if the corresponding $\Delta S_\phi$ is positive, or else with probability $e^{-\beta \Delta S_\phi}$. If we take $\beta=1/(k_BT)$ large enough (implying a sufficiently low temperature), the simulation will sample in an effective way all the states with $\beta S \ll 1$; it is possible to monitor the process by recording the maximum value of $S$ attained, or better with a histogram or time-series of all values of $S$. In suitable units, we shall therefore satisfy the condition $S_\phi \ll \hbar$, so we are effectively sampling the configurations with an action very close to the minimum, and we can compute in this sample averages of quantities like $\langle \phi_k \rangle$, $\langle (\phi_k-1)^2 \rangle$, obtaining respectively results $\simeq 1$ and $\ll 1$.

\subsection{Back to gravity: $S_E$ is not positive, then sample with $S_E^2$ or $|S_E|$}

Now let us go through the same steps for the gravitational field $A$ and its path integral (\ref{PI-A}). The classical vacuum value and the initial conditions are the same: $A_h=1$, $h=1,\ldots,N$. The action $S_E$, however, is not positive-defined. In the Lorentzian path integral with the Wigner paths, with small amplitude $a$, we can see by monitoring $S_E$ that it changes in sign, but still it remains small in absolute value and the results for $\langle A \rangle_k$ in this limit are similar to those for $\langle \phi_k \rangle$. When $a$ is increased, the positive and negative fluctuations in $S_E$ increase, adding to the noise coming from the factor $e^{iS/\hbar}$. It would not make sense to pass, like for $\phi$, to a Metropolis simulation based on $S_E$ in order to obtain the ``thermal sampling'' of the region with $S_E \ll \hbar$. We can do, however, the Metropolis sampling with $S_E^2$, using the probability $e^{-\beta^2 \Delta S_E^2}$. As with the $\phi$-field, we can check that during the runs the values of $S_E$ remain very close to the minimum and we can compute the averages of $\langle A_k \rangle$, $\langle (A_k-1)^2 \rangle$. The results are given and discussed in Sect.\ \ref{res}, and are very different from those for $\phi$! Large deviations from the classical values $A_k=1$ are observed. 

Is is important, at this point, to consider the following natural objection: minimizing $S^2_E$ is not the same as minimizing $S_E$, because $\Delta (S^2_E)=2S_E \Delta S_E$, so that a classical theory based on the action $S_E^2$ has more solutions than one based on $S_E$. Actually, the additional classical solutions are just the zero modes discussed previously, for which $S_E=0$. In order to reply to this objection we note that:
 
(1) The configurations with $S_E \ll \hbar$ obtained from the Metropolis algorithm with $S^2_E$ are very different from the classical zero modes of $S_E$. Therefore such configurations cannot originate from the spurious minimum of $S^2_E$ at $S_E=0$ and it appears that, ironically, the only use of the classical zero mode solutions is for making sure that they are irrelevant (probably due to entropic or phase-space reasons). (2) As a further check, we have run the algorithm also with a probability based on $|S_E|$. Results are very similar (see Sect.\ \ref{alt}). Although the absolute value looks a bit cumbersome and was not our first choice, it has the advantage that no spurious stationary points are introduced.

\section{Classical, exact zero modes}
\label{cla}

In this Section we give a short review of our previous work on classical, exact zero modes of the Einstein action. These are metrics for which $S_E=0$ ($S_E$ is defined in eq.\ (\ref{eq1})), or in other words metrics whose scalar curvature has vanishing integral.
 
In perturbation theory, one can obtain such zero modes defining, at a purely mathematical level, some suitable unphysical sources and solving the linearized Einstein equation for these sources. The Einstein equations with source $T_{\mu \nu}$ obtained from the Einstein action are, as usual
\begin{equation}
R_{\mu \nu}-\frac{1}{2}g_{\mu \nu}R=-\frac{8\pi G}{c^4} T_{\mu \nu}
\end{equation}
By contraction with $g^{\mu \nu}$ one obtains
\begin{equation}
R=\frac{8\pi G}{c^4} g^{\mu \nu}T_{\mu \nu}=\frac{8\pi G}{c^4}T_\rho^\rho
\end{equation}
Therefore if we define a source such that
\begin{equation}
\int d^4x \sqrt{g(x)}T_\rho^\rho(x)=0
\end{equation}
its field will be a zero mode. If the source is static, we can suppose for simplicity that only the $T_{00}$ component is non vanishing. The factor $\sqrt{g}g_{00}$ can be expressed in function of $T_{00}$ through the Feynman propagator and the conclusion is that
\begin{equation}
\sqrt{g(x)}g_{00}(x)=1+o(G^2)
\end{equation}
Therefore the action can simply be rewritten, for that source, as
\begin{equation}
S_E=-\frac{1}{2} \int d^4x T_{00}(x) +o(G^2)
\end{equation}
and in order to have $S_E=0$ at this level of approximation it is sufficient to choose an unphysical source $T_{00}(x)$ of the form of a mass dipole, or one with spherical symmetry, with two concentric mass shells of opposite sign \cite{modanese2000large}.
 
Turning to the strong-field case, in \cite{modanese2007vacuum} we used a virtual source method to search for zero modes of the action. In the
context of wormhole physics \cite{lobo2004fundamental} this method is also called ``reverse solution of the Einstein
equations''; it consists in finding a source which generates a metric with some desired features. 
The goal here would be to find zero modes with curvature polarization, namely having two regions with positive and negative scalar curvature such that the total integral of $\sqrt{g}R$ is zero. However, this turns out to be impossible, because the source must also satisfy independently the Bianchi identities, and the two conditions are found to be incompatible beyond the linear approximation.

Still there exist in the strong-field case a set of static and spherically symmetric zero modes without polarization. In fact the condition that the integrand of the action (\ref{eq8}) be identically zero can be written as an ordinary differential equation in the unknown $A(r)$:
\begin{align}
	\frac{rA'(r)}{A^2(r)}+1-\frac{1}{A(r)}=0
\end{align}
It is straightforward to check that the solutions of this equation are scale invariant, i.e., if $A(r)$ is a solution, then also $A(kr)$ is a solution, for any positive $k$. The non-singular solution such that $A(r) \to 1$ when $r \to \infty$ is
\begin{align}
	A(r)=\frac{r}{r+\mu}, \qquad \mu \geq 0
	\label{zm1}
\end{align}
One can check that this solution has for large $r$ the same form as a Schwarzschild metric with mass $-\mu/G$. It is regular everywhere, monotonically increasing from $A(0)=0$, and equal to $\frac{1}{2}$ when $r=\mu$.

Finally one can define small periodic deformations of the zero modes (\ref{zm1}) which still have null integral, although they do not have $R=0$ everywhere (see details in \cite{modanese2007vacuum}).

In conclusion, there exist for the Einstein action weak classical zero modes of the polarized kind and strong field zero modes which are not polarized. We have discussed analytically the possible role of these modes in the path integral in \cite{modanese2012anomalous}. In the next section we shall see what happens in strong field lattice simulations.

\section{Results of the quantum simulations}
\label{res}

\begin{figure}[t]
\begin{center}
\includegraphics[width=10cm,height=6cm]{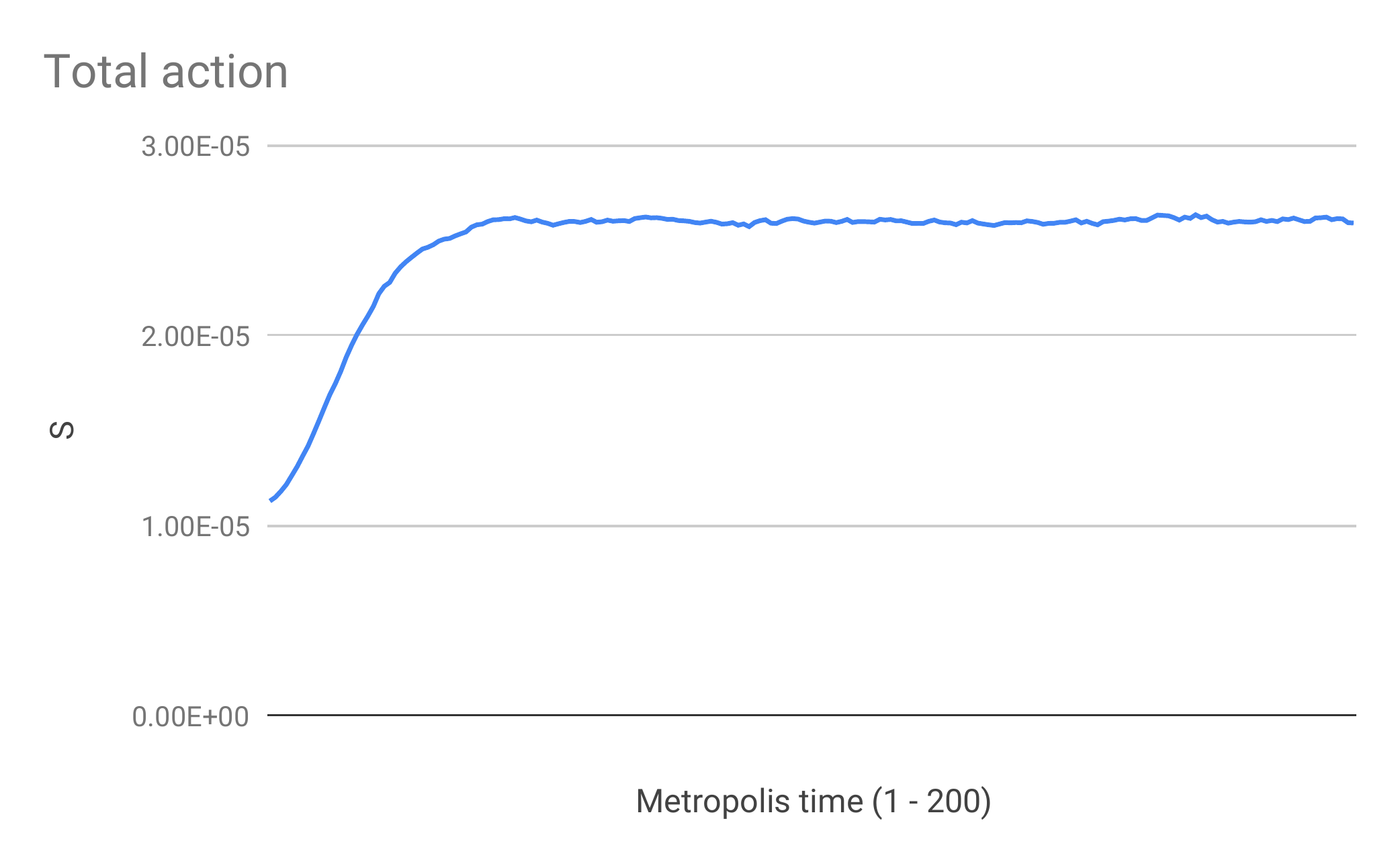}
\caption{Values of the action measured during a typical Metropolis run with inverse temperature $\beta=10^{13}$. An equilibrium value of the order of $10^{-5}$ is attained (all quantities in units such that $\hbar=c=G=1$; interval length $L=1$, number of sub-intervals $N=100$).
} 
\label{fig1}
\end{center}  
\end{figure}

The best choice of the parameters in the Metropolis simulation for attaining thermal equilibrium is found to be
 $\varepsilon=10^{-6}$, $\beta=10^{13}$. With this choice the total action $S_E$ reaches quickly a small positive value of the order of $10^{-5}$ (Fig.\ \ref{fig1}). A typical run includes 200 time steps, each with $5\cdot 10^6$ local ``spin flips'' of the variables $A_h$. The average values of $A_h$ are computed in the second half of the run, after equilibration. Typical figures for other averages of interest are
$\langle S \rangle = 2.6 \cdot 10^{-5}$ and $\langle e^{-\beta^2 \Delta (S^2)} \rangle = 0.98$
(acceptance ratio for changes with $\Delta (S^2) >0$). An example of the C code used is appended to the TeX source of this preprint.

\begin{figure}[t]
\begin{center}
\includegraphics[width=10cm,height=6cm]{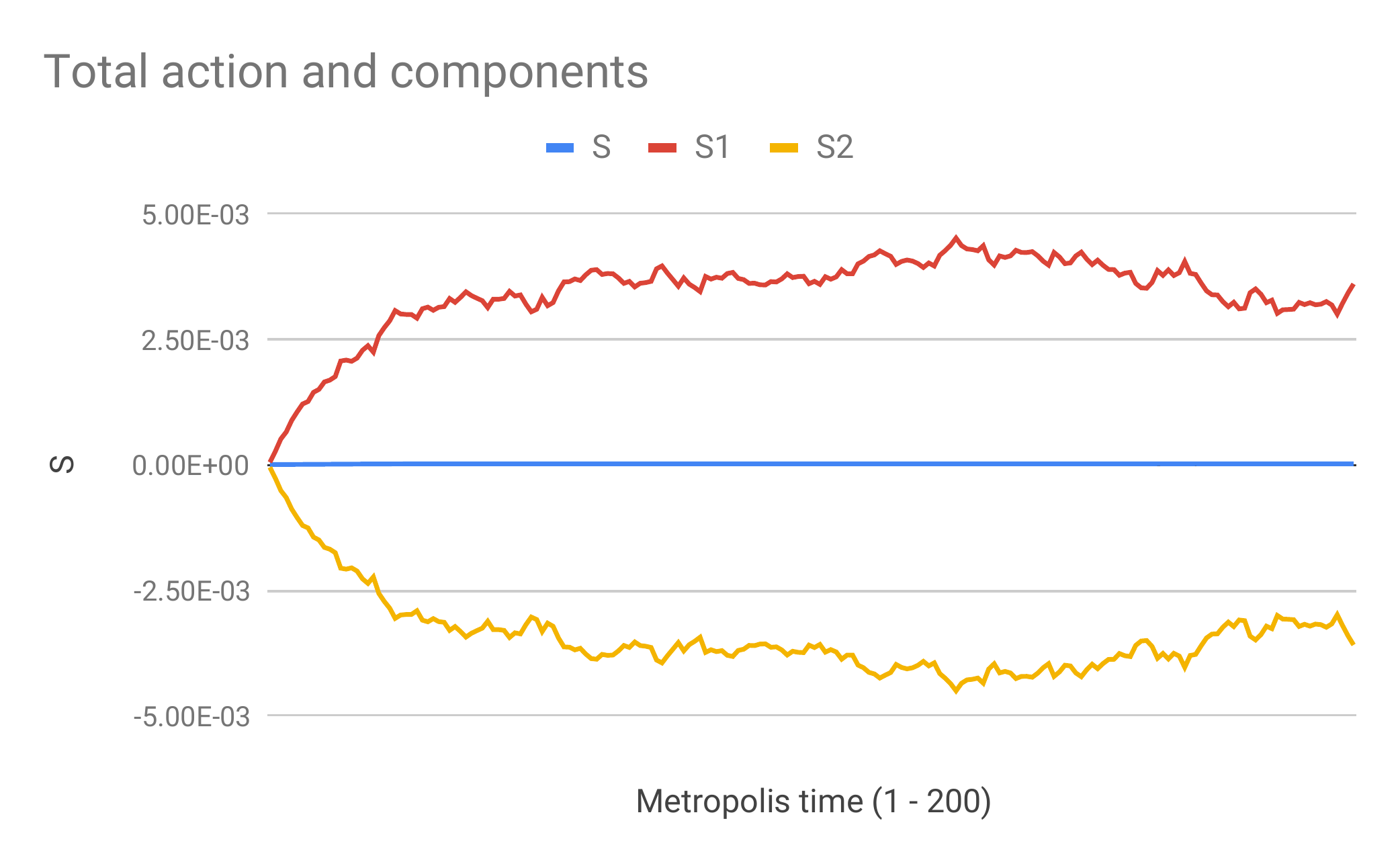}
\caption{Components $S_1$ and $S_2$ of the action for the same run of Fig.\ \ref{fig1}. $S_1$ is given by the sum from 1 to $\frac{N}{2}$, i.e.\ on the left half of the interval. $S_2$ is given by the sum from $\frac{N}{2}+1$ to $N$, i.e.\ on the right half. Both $S_1$ and $S_2$ are much larger, in absolute value, than the total action. Since the sign of $R$ is opposite to that of the local $S$, we conclude that the inner part of the metric has negative curvature and the outer shell has positive curvature. The total integral of the curvature is small and negative.} 
\label{fig2}
\end{center}  
\end{figure}

\begin{figure}[t]
\begin{center}
\includegraphics[width=10cm,height=6cm]{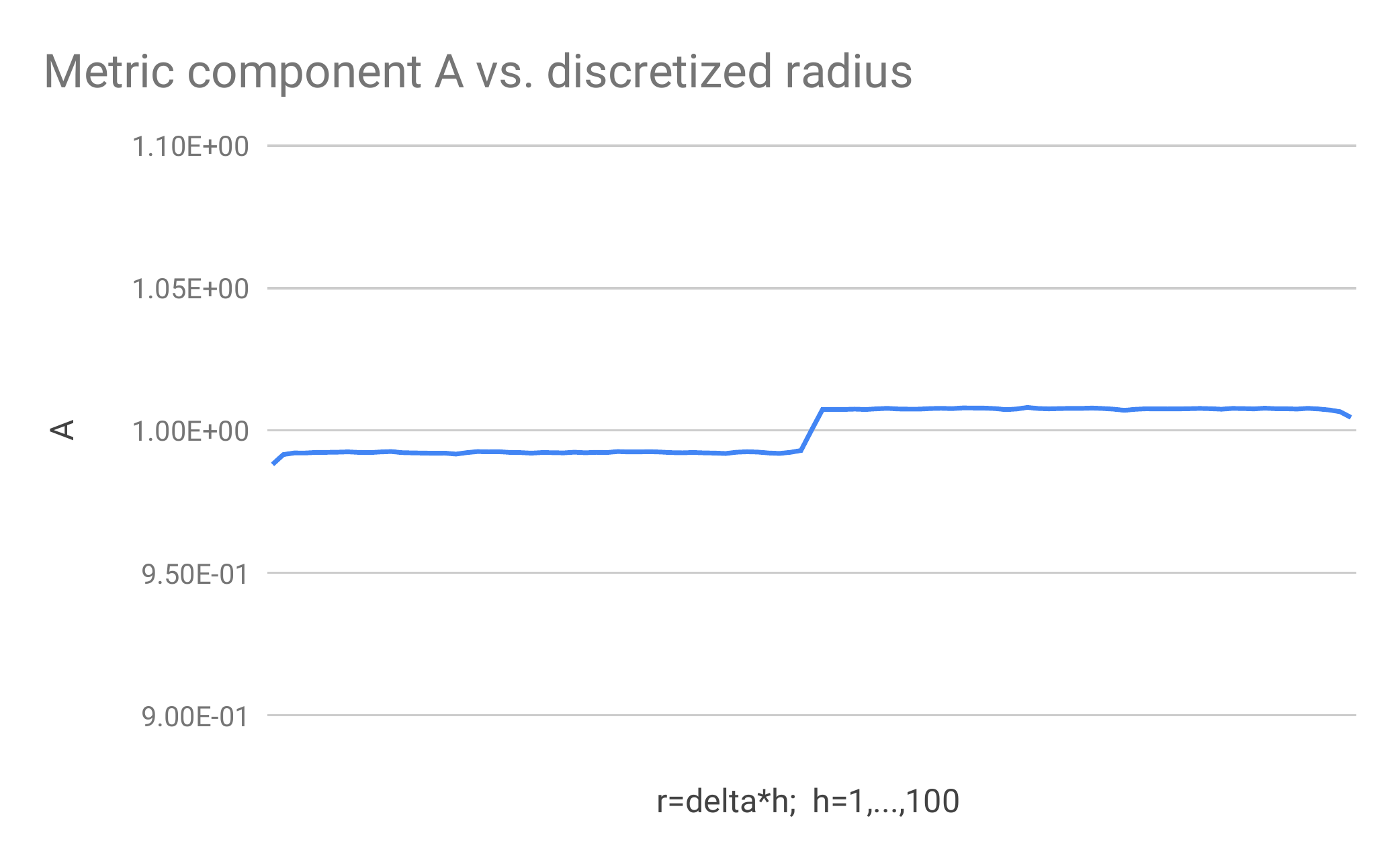}
\caption{Metric component $A=g_{rr}$ as a function of the discretized radius $r=\delta h$ ($\delta=L/N$ is the lattice cutoff). The condition at the right boundary is $A=1$.
} 
\label{fig3}
\end{center}  
\end{figure}

Fig.\ \ref{fig2} shows that the component $S_1$ of the action (the one given by the sum in the interval $0<r<L/2$, i.e., with index $h$ from 1 to $N/2$) is positive, and vice-versa for the component $S_2$. From Fig.\ \ref{fig3} we see that in the same interval the metric $A(r)$ is approximately constant, say $A \simeq 1-\chi$ ($0<\chi \ll 1$), and conversely in the interval between $L/2$ and $L$ we have $A \simeq 1+\chi$.

In the expression (\ref{eq8}) for the action we observe that if $A$ is constant in an interval, then the integrand reduces to $(1-\frac{1}{A})$. By replacing $A \simeq 1 \pm \chi$, we find that the integrand is approximately equal to $-\chi$ in the left interval and to $+\chi$ in the right interval, i.e., it is opposite to the values of $S_1$ and $S_2$ obtained numerically.

\begin{figure}[t]
\begin{center}
\includegraphics[width=10cm,height=6cm]{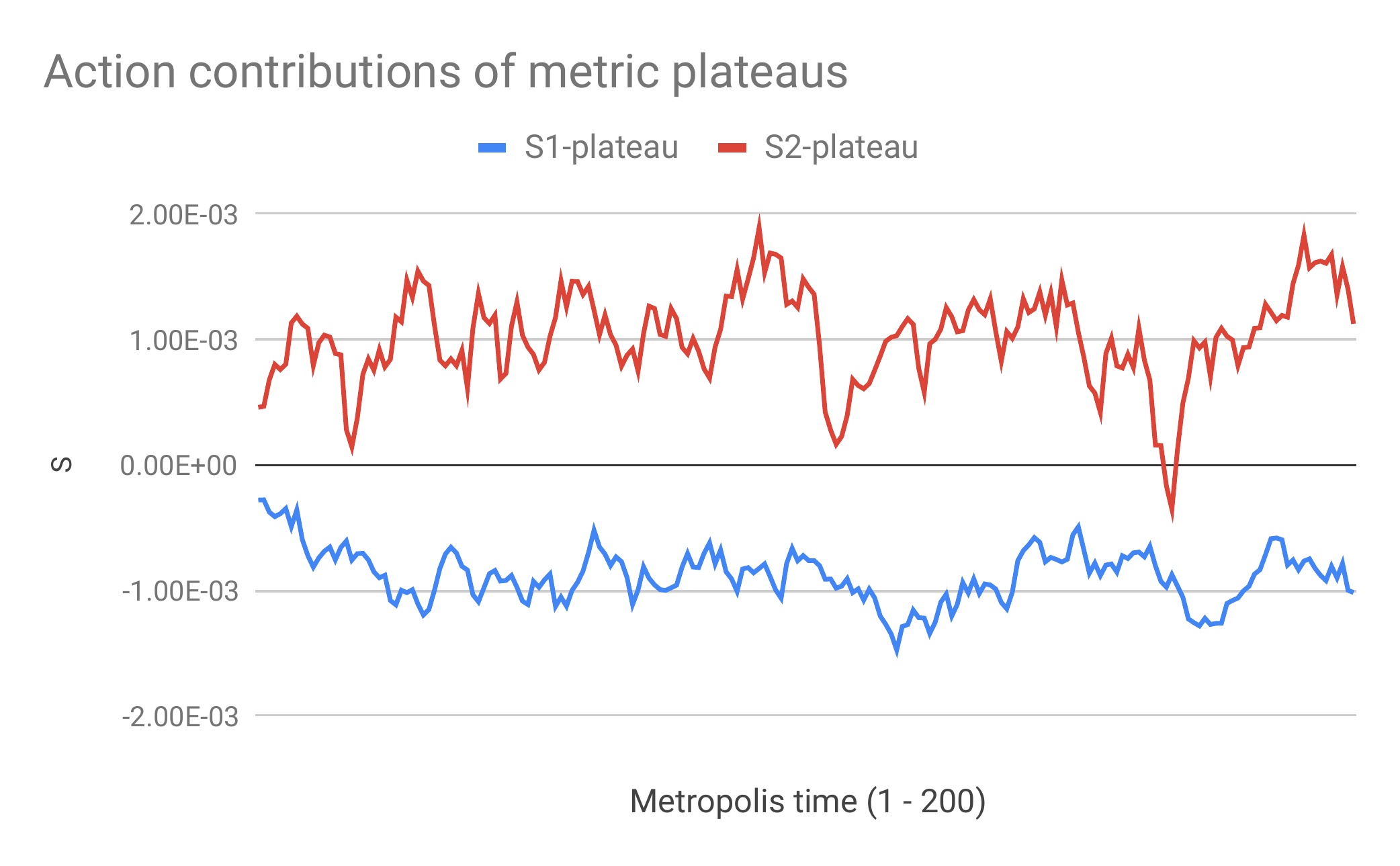}
\caption{Action contributions from two metric ``plateaus'' where the metric is approximately constant.
The contribution $S_{1-plateau}$ comes from the inner region with $30 \le h \le 40$. The contribution $S_{2-plateau}$ comes from the outer region with $60 \le h \le 70$. Note that the sign of $S_{1-plateau}$ is opposite to the sign of $S_1$, and the same holds for $S_{2-plateau}$ and $S_2$.
} 
\label{fig4}
\end{center}  
\end{figure}

This means that the main contributions to the action do not come from the ``plateaus'' with constant metric (Fig.\ \ref{fig4}), but from the ``steps'' at $r=0$, $r=L/2$ and $r=L$, where $R$ cannot be estimated as easily as in the plateaus. It is mainly the right combination of steps which makes the action vanish. Note that these steps build up spontaneously and in a reproducible way in the thermalization algorithm from billions of random flips of the discretized variables $A_h$.

The occurrence of one step exactly at the middle of the interval $(0,L)$ is most likely an artifact of the discretization (and remember that physically $L$ must tend to infinity). Still a physical phenomenon of polarization between regions with $R<0$ and regions with $R>0$ clearly emerges from these results. Being $r$ the radial coordinate of a spherical configuration, the two polarization regions are actually one sphere and one spherical shell around it. Since $R$ has opposite sign to the integrand of the action, we conclude that in this class of vacuum fluctuations the inner region has $R<0$ and the outer region has $R>0$.

For comparison with the exact, classical zero modes of Sect.\ \ref{cla} (with $S_E=0$) we recall that polarized zero modes are obtained in the linearized approximation, but they do not survive in strong field. Polarized zero modes in strong field appear thus to be present, and entropically dominant, only under the weaker condition $S_E \ll \hbar$.

\subsection{Results with transition probability $\exp(-\beta\Delta |S_E|)$}
\label{alt}

As mentioned in Sect.\ \ref{sim}, simulations have also been performed with a transition probability based on $|S_E|$, as an alternative to $S_E^2$. With parameters $\varepsilon=10^{-6}$, $\beta=10^9$ one obtains results very similar to those those with probability $\exp(-\beta^2\Delta S_E^2)$ and parameters $\varepsilon=10^{-6}$, $\beta=10^{13}$ (Figs.\ \ref{fig1}, \ref{fig2}; the relation between the values of $\beta$ in the two cases is not obvious, since $\varepsilon$ is also involved). Again we observe in the simulation a positive action which increases until it reaches a value of the order of $10^{-5}$ and then stays approximately constant, while the polarized components $S_1$ and $S_2$ undergo small oscillations around values of the order of $10^{-3}$.

\begin{figure}[t]
\begin{center}
\includegraphics[width=10cm,height=6cm]{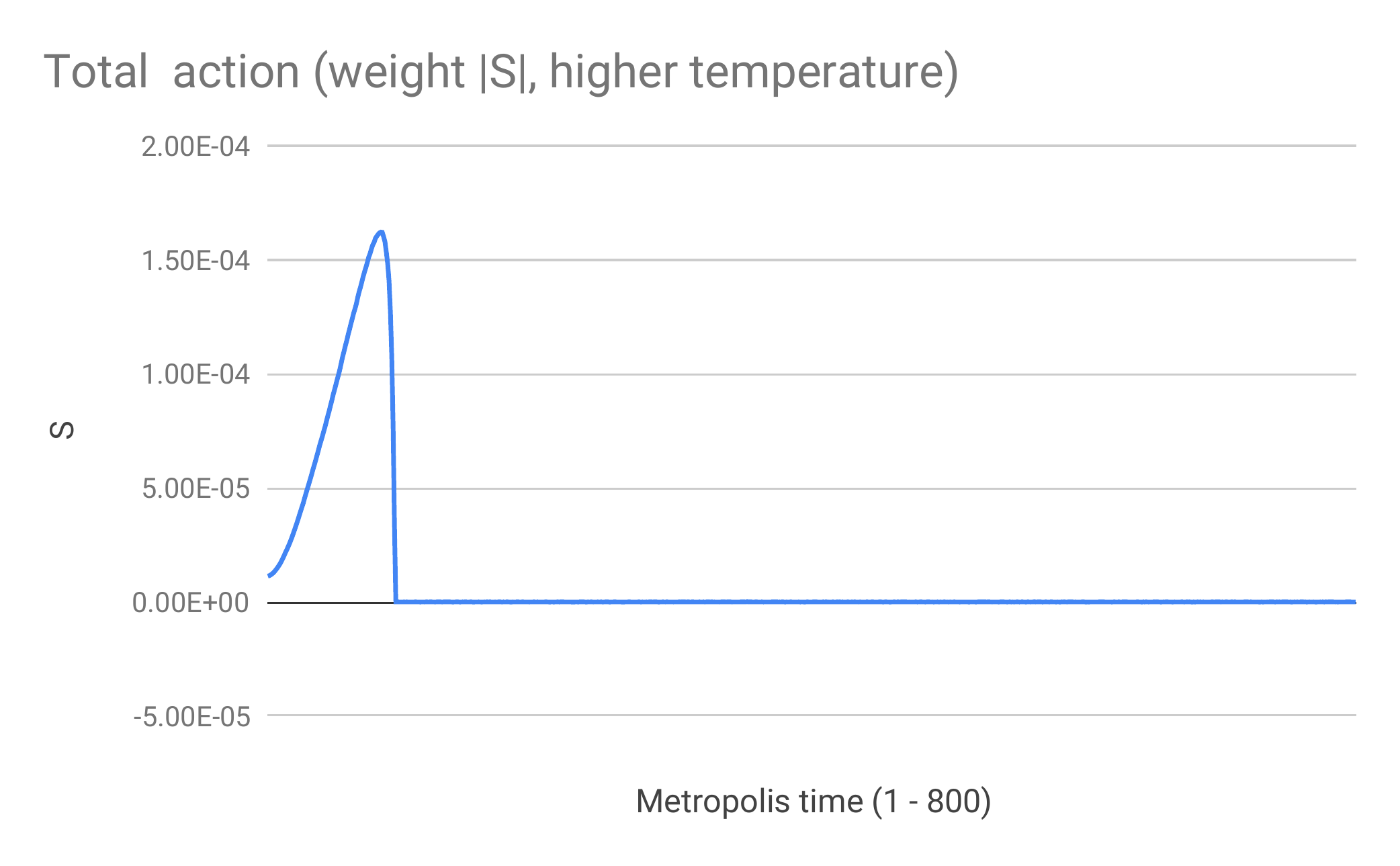}
\caption{Total action with probability $e^{-\beta|S|}$ and inverse temperature $\beta=10^8$. It reaches a positive maximum value of the order of $10^{-4}$ and then decreases to $10^{-9}$. 
} 
\label{fig5}
\end{center}  
\end{figure}

\begin{figure}[t]
\begin{center}
\includegraphics[width=10cm,height=6cm]{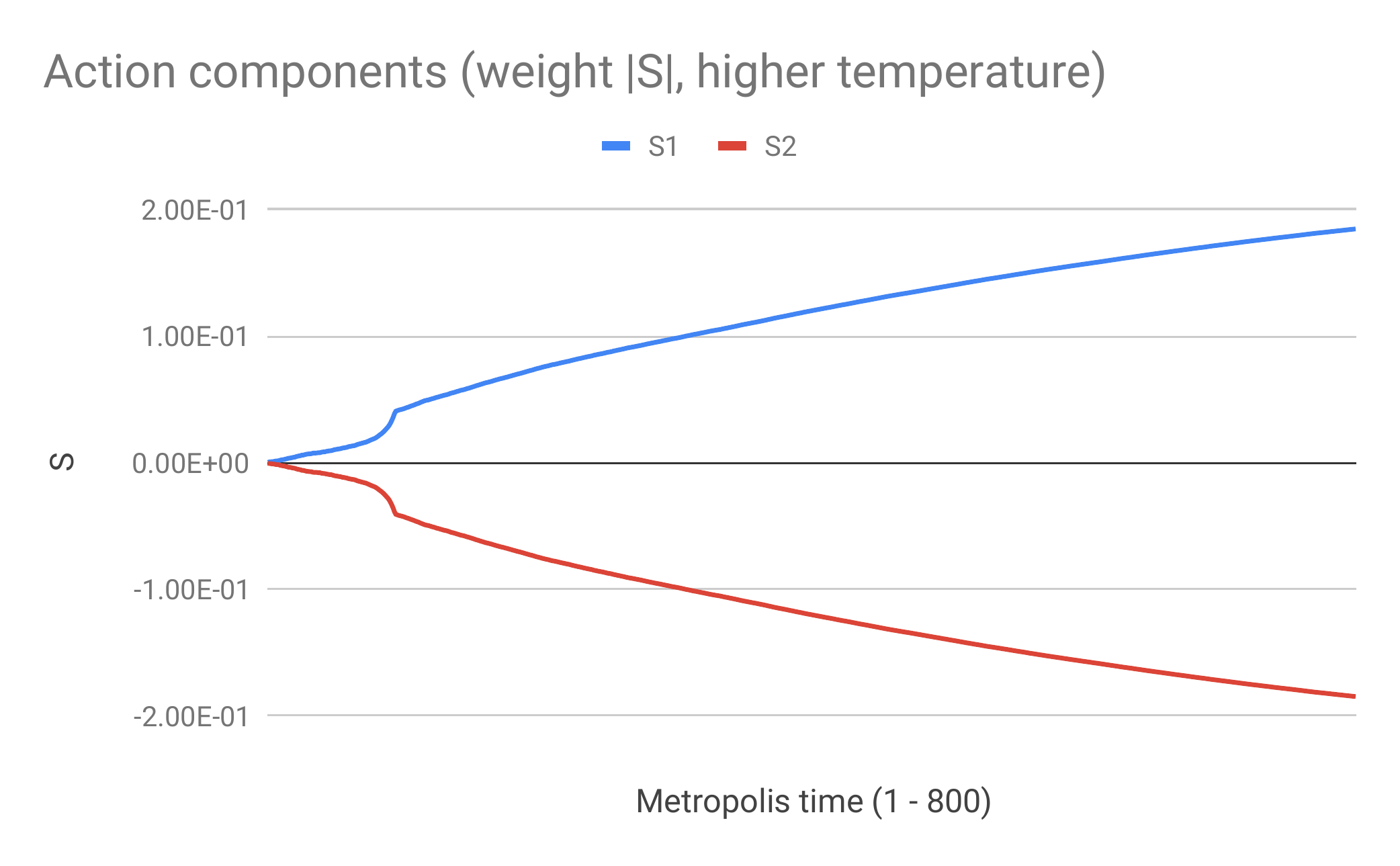}
\caption{Components $S_1$ and $S_2$ of the action for the same run of Fig.\ \ref{fig5}. They are about 8 magnitude orders larger than the total action.
} 
\label{fig6}
\end{center}  
\end{figure}

With a tenfold increase in ``temperature'' ($\beta=10^8$) some different field configurations are obtained, which are quite interesting for our purposes (Fig.\ \ref{fig5}). After a moderate growth to the magnitude $10^{-4}$ as before, the action returns very close to zero ($|S| \sim 10^{-9}$, with oscillations of both positive and negative sign); in the meanwhile, the polarized components $S_1$ and $S_2$ (Fig.\ \ref{fig6}) are larger by approximately 8 magnitude orders, apparently resulting from a very efficient compensation mechanism. Unlike in the previous case, however, the components $S_1$ and $S_2$ do not appear to reach an equilibrium value but continue to grow.

\begin{figure}[t]
\begin{center}
\includegraphics[width=10cm,height=6cm]{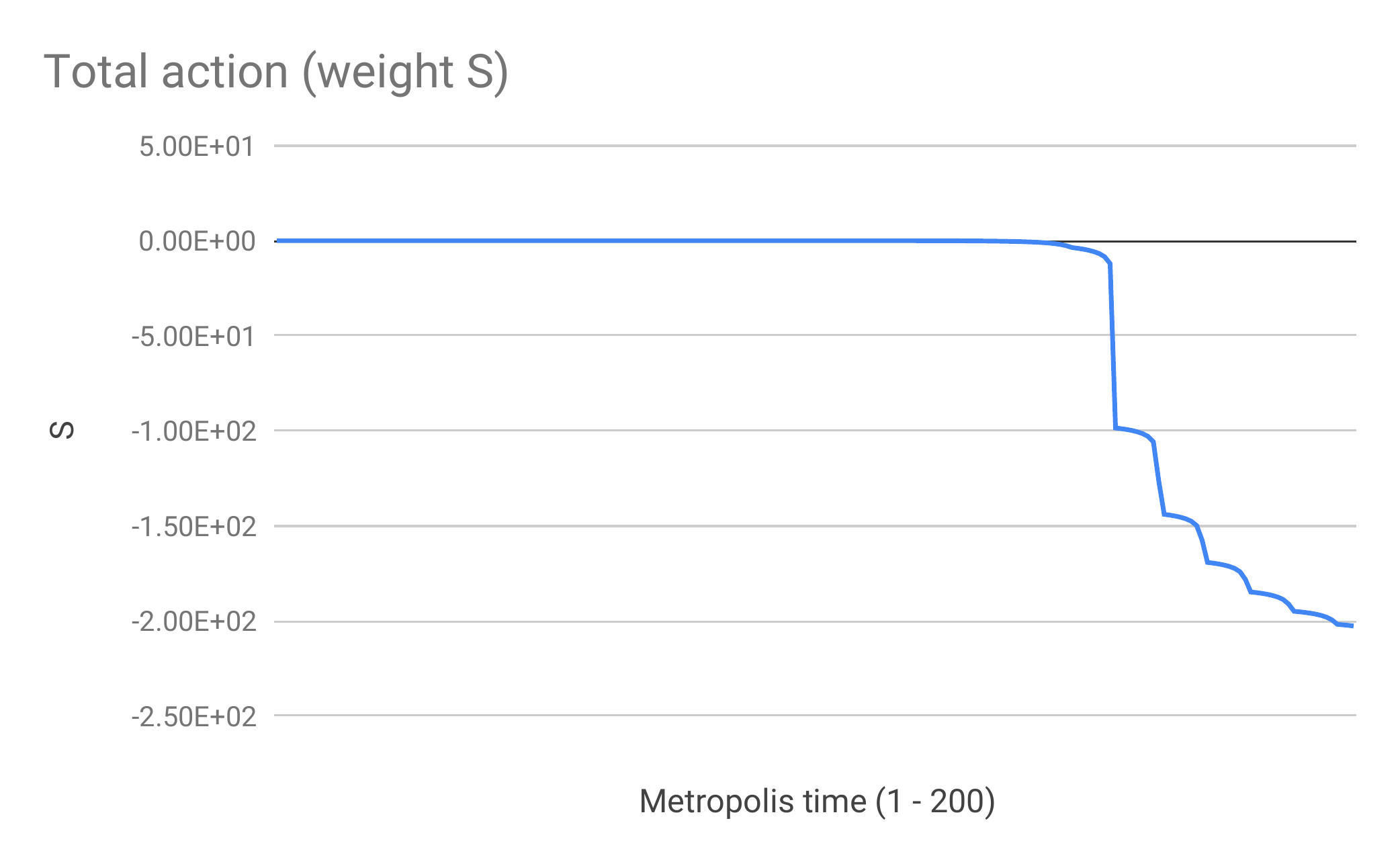}
\caption{Total action with probability $e^{-\beta S}$ and inverse temperature $\beta=10^8$. The system is clearly unstable with respect to negative fluctuations of the action.
} 
\label{fig7}
\end{center}  
\end{figure}

Finally, it is interesting to note that if we make simulations simply with probability $\exp(-\beta S_E)$, disregarding the problem of the indefinite sign of $S_E$, then with very low temperature ($\beta=10^9$ or more) one still obtains a positive growth of $S$ to $\sim 10^{-5}$ followed by stabilization, but going to smaller $\beta$ one observes a collapse of the action to large negative values (Fig.\ \ref{fig7}). The corresponding metric shows large deviations from flat space.

\section{Comparison with other approaches}
\label{com}

A fundamental reference point for the understanding of non-perturbative quantum gravity (QG) in the path integral approach is in our opinion the work of H. Hamber, contained in several seminal papers starting from 1984 and summarized in the book \cite{hamber2008quantum} and in a few review articles. Among the latter we can mention one from 2009 \cite{hamber2009quantum} and the very recent article appeared in this same Special Issue \cite{hamber2019vacuum}, containing recent results on the quantum condensate which appears to be present in the vacuum state of QG.

Hamber regards non-perturbative quantum chromodynamics (QCD) as a possible model for QG, though with some important differences. In particular, in QG the cutoff or lattice spacing $l_0$ of the discretized theory remains implicitly present in the physical value of $G$, and there exists another dynamically generated scale $\xi$, the correlation length of the curvature. A meaningful lattice limit is obtained when $\xi \gg l_0$. Although the magnitude of $\xi$ is not directly related to the value of $G$, it signals how close the bare $G$ is to the fixed point value $G_c$.

These statements refer to the renormalization group analysis and to the scaling analysis in dependence on the cutoff that is necessary in order to find the correct continuum limit of lattice field theories \cite{zinn1996quantum}, and which in this paper has not yet been  performed, except for noting the scale invariance of the classical zero modes.

Actually, in our approach we do not admit the very general scale invariance considered by Hamber, according to which the metric can undergo transformations of the form $g'_{\mu \nu}=\omega g_{\mu \nu}$. We assume that the metric is always asymptotically equal to the flat metric $\eta_{\mu \nu}$, and so in our simulations the component $A(r)=g_{rr}(r)$ has always boundary condition $A=1$ at large $r$. This is also because we do not have a cosmological term in the action. Hamber, on the contrary, starts from an action containing a bare cosmological constant $\lambda_0$ and looks for a mechanism of dynamical generation of flat spacetime. He finds that the system reaches a stable ground state only if $\lambda_0>0$; our (limited) results are compatible with this fact, since the vacuum fluctuations we find have average negative scalar curvature, such to compensate a positive $\lambda_0$ and bring the system closer to flat space.

Further enlarging the picture, it should be mentioned that according to Hamber the large-scale average $\left\langle R \right\rangle$ in pure QG is not zero but equal to $1/\xi^2$. Therefore $\xi$ is related to a gravitational ``condensate'', physically represented by the observed cosmological constant. This establishes a very interesting link between QG and cosmology \cite{hamber2019vacuum}.

Other possible approaches to the problem of the true ground state in quantum gravity are those of Preparata and collaborators \cite{preparata2000gas} and of Garattini \cite{garattini2002spacetime}. Both are based on a Hamiltonian formulation. Classical wormhole metrics with spherical symmetry are taken as candidates for the ground state (with lower energy than flat space) and the spectrum of the quantum excitations with respect to this background is computed. In Ref.\ \cite{preparata2000gas} a ``gas of wormholes'' is then considered, as macroscopic limit, and it is shown that it yields on the average a flat metric. In Ref.\ \cite{garattini2002spacetime} it is further argued that a coherent superposition of wormholes, and not the single wormholes, is privileged with respect to flat space. In both approaches, strong quantum fluctuations are found to occur only at the Planck scale. In contrast, the action zero modes of \cite{modanese2007vacuum} occur at any scale.

\section{Conclusions and outlook}
\label{con}

The difficulties encountered in the formulation and numerical simulation of lattice quantum gravity are manifold. This is certainly to be expected, since in quantum gravity the spacetime lattice itself takes part to the dynamics. In this general context our present contribution is quite limited, also because we consider only time-independent metrics with spherical symmetry. Nevertheless, we have been able to single out some phenomena which occur in the nonperturbative lattice dynamics and can perhaps be present under less specific circumstances. Such phenomena are quite intuitive from the physical point of view. The idea of zero modes of the action with polarization into regions of positive and negative scalar curvature, already explored at the perturbative level, is confirmed under strong field conditions.

More in detail, we can summarize our previous and present results as follows:
\begin{enumerate}
\item Exact zero modes ($S_E=0$) with curvature polarization exist in the weak field approximation.
\item Exact zero modes without curvature polarization exist also in strong field, but
\item Quantum zero modes ($S_E \ll \hbar$) of any strength are predominantly of the polarized kind.
\end{enumerate}

The ``trick'' of weighing the configurations in the path integral with $e^{-\beta^2 S_E^2}$ (or $e^{-\beta |S_E|}$) instead of $e^{-\beta S_E}$, in order to avoid the problem of the indefinite sign of $S_E$, does not seem to be a weak point of the argument. What is still missing is a scaling analysis of the results, which will be presented in a forthcoming work.

A further extension of the results could be obtained by varying in the path integral not only the metric component $g_{rr}$, but also $g_{00}$. From the analytical point of view this looks intractable, but in the numerical code it should not make a big difference, except for doubling the number of degrees of freedom and complicating the expression of the discretized action.

Finally, an open question is what average physical quantities can be defined and measured in this kind of simulations, besides the local curvature and metric, which show the occurrence of the polarization phenomenon but not its possible consequences.

\bibliographystyle{unsrt}
\bibliography{QG}

\end{document}